\documentstyle[sprocl,epsfig]{article}

\def\NIMA{{\em Nucl. Instrum. Methods} A}

\def\be{\begin{equation}}
\def\ee{\end{equation}}
\def\bea{\begin{eqnarray}}
\def\eea{\end{eqnarray}}

\begin{document}

\def\thefootnote{\fnsymbol{footnote}}

\hfill {\tt LC Note: LC-TOOL-2004-020}

\hfill {\tt arXiv:physics/0409039}

\vspace{33pt}

\title{CALORIMETER CLUSTERING WITH MINIMAL SPANNING TREES\\}

\author{GEORGIOS MAVROMANOLAKIS~\footnote[1]{~email: {\tt gmavroma@hep.phy.cam.ac.uk} or {\tt gmavroma@mail.cern.ch}} 
}

\address{Cavendish Laboratory, University of Cambridge\\
Madingley Road, Cambridge CB3 0HE, U.K.}

\maketitle\abstracts
{
We present a top-down approach to calorimeter clustering. 
An algorithm based on minimal spanning tree theory is described briefly.
}

\vfill
\begin{center}
Proceedings of the International Conference on Linear Colliders \\
LCWS 2004, Paris, 19-23 April 2004
\end{center}
\vfill

\thispagestyle{empty}

\clearpage

\setcounter{page}{1}

\title{CALORIMETER CLUSTERING WITH MINIMAL SPANNING TREES\\}

\author{GEORGIOS MAVROMANOLAKIS}

\address{Cavendish Laboratory, University of Cambridge, Cambridge CB3 0HE, U.K.}

\maketitle\abstracts
{
We present a top-down approach to calorimeter clustering. 
An algorithm based on minimal spanning tree theory is described briefly.
}

\section{Introduction}

Clustering calorimeter hits is a complex pattern recognition problem 
with complexity depending on event type, energy and detector design.
A successful clustering algorithm must be characterised by high efficiency 
and speed to cope with and to exploit the high granularity design forseen for 
both electromagnetic and hadronic calorimeters in a Future Linear Collider 
experiment.
In the following we describe a top-down or divisive hierarchical clustering approach
where the entire set of hits is first considered to be a single cluster, the minimal 
spanning tree, which is then broken down into smaller clusters.

\section{Clustering With Minimal Spanning Trees}

\begin{figure}[tbp]
\begin{center}
\vspace{-10pt}
\begin{tabular}{ccc}
\mbox{\epsfig{file=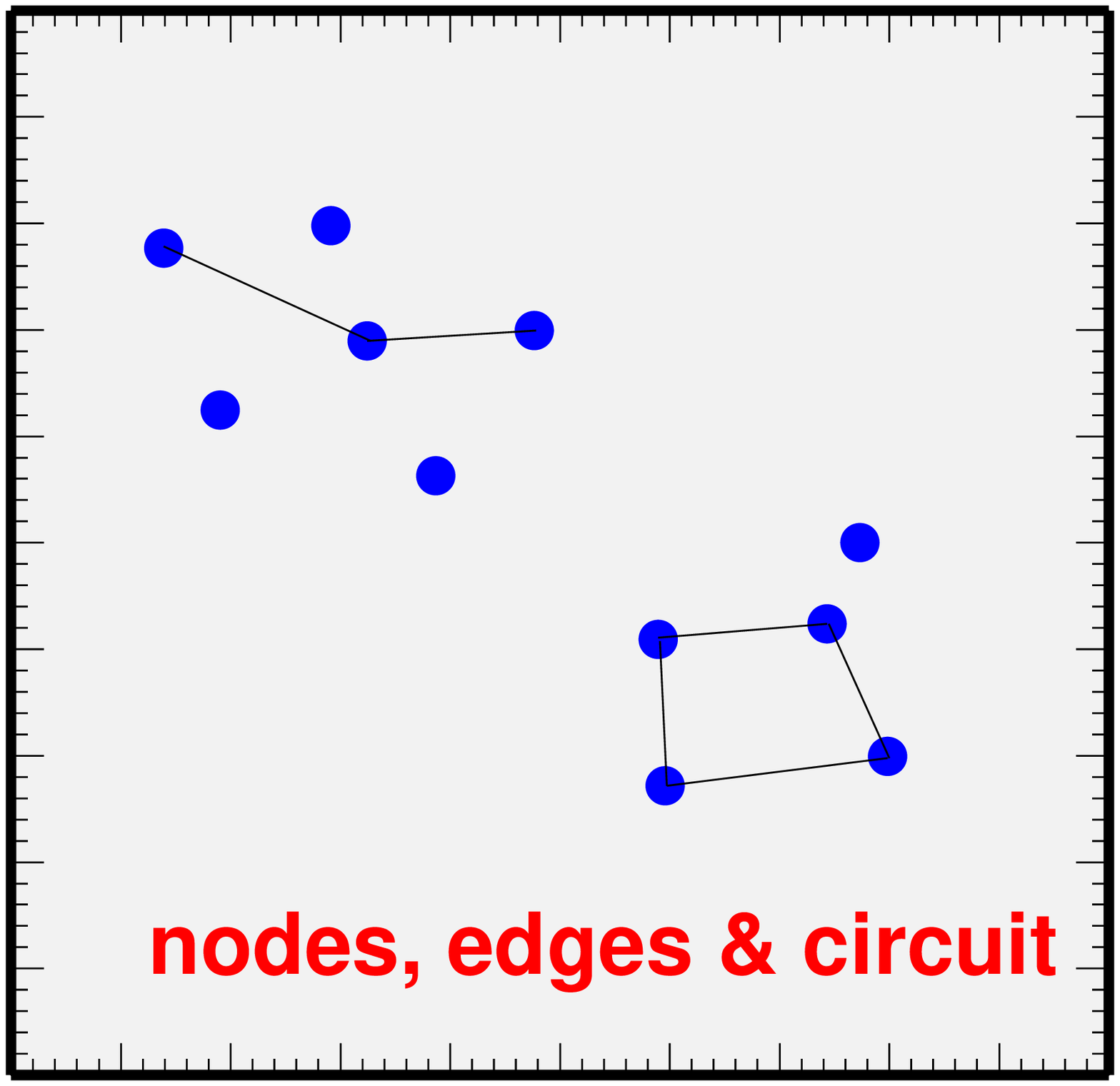,width=3.3cm}}&
\mbox{\epsfig{file=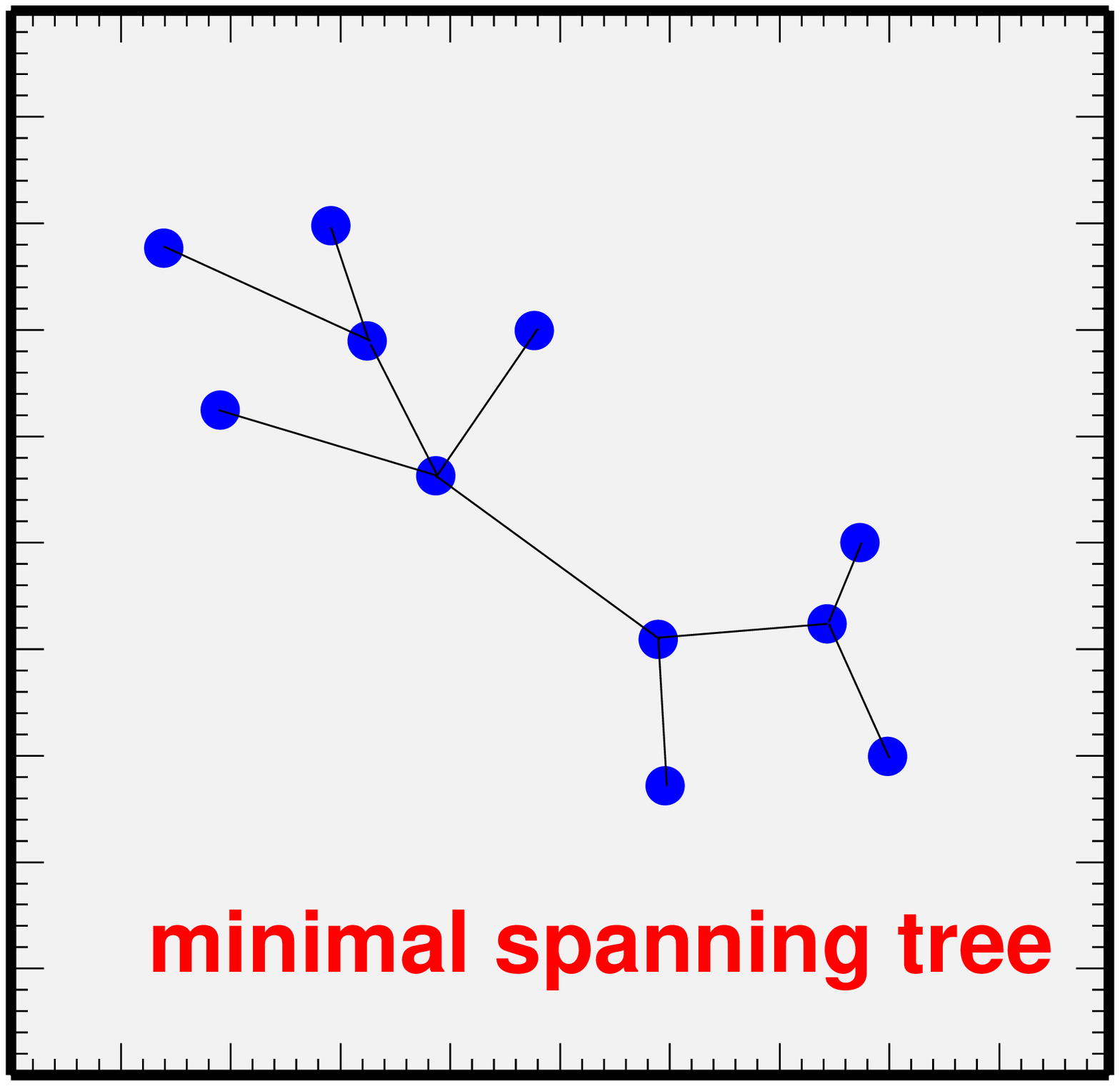 ,width=3.3cm}}&
\mbox{\epsfig{file=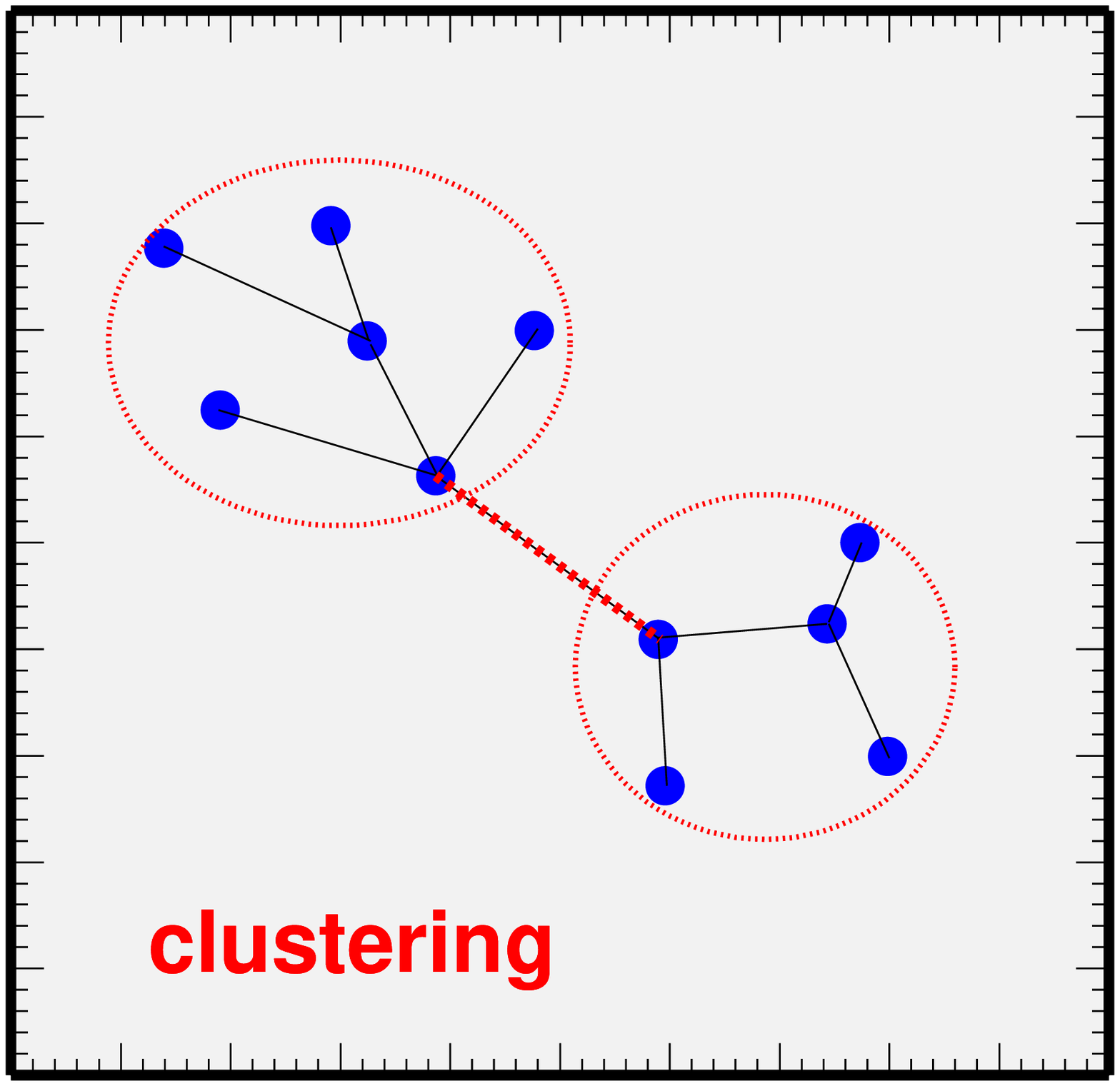 ,width=3.3cm}} \vspace{-18pt}\\
\end{tabular}
\end{center}
\caption{Illustration of terms and concepts discussed, nodes, edges and circuit, 
minimal spanning tree, single linkage cluster analysis.}
\label{fig:one}
\end{figure}

Given a set of nodes in a configuration space and a metric to 
assign distance cost or weight to each edge connecting a pair of nodes, we define 
the minimal spanning tree as the tree which contains all nodes with no circuits
and of which the sum of weights of its edges is minimum (see Fig.~\ref{fig:one}). 
A minimal spanning tree is unique for the given set of nodes and the chosen 
metric, it is deterministic {\em i.e.} it has no dependency on random choices 
of nodes during construction, and it is invariant under similarity transformations 
that preserve the monotony of the metric [\ref{zahn}]. 
First developed and applied to problems related to efficient design of 
networks [\ref{prim}], minimal spanning trees are well studied mathematical objects 
and there is a solid base of theorems which relate them to efficient clustering 
as well [\ref{zahn}]. Applications to high energy physics 
can be found in [\ref{saoulidou}]. 

\begin{figure}[tbp]
\vspace{-1.5mm}
\begin{center}
\begin{tabular}{ccc}
\mbox{\epsfig{file=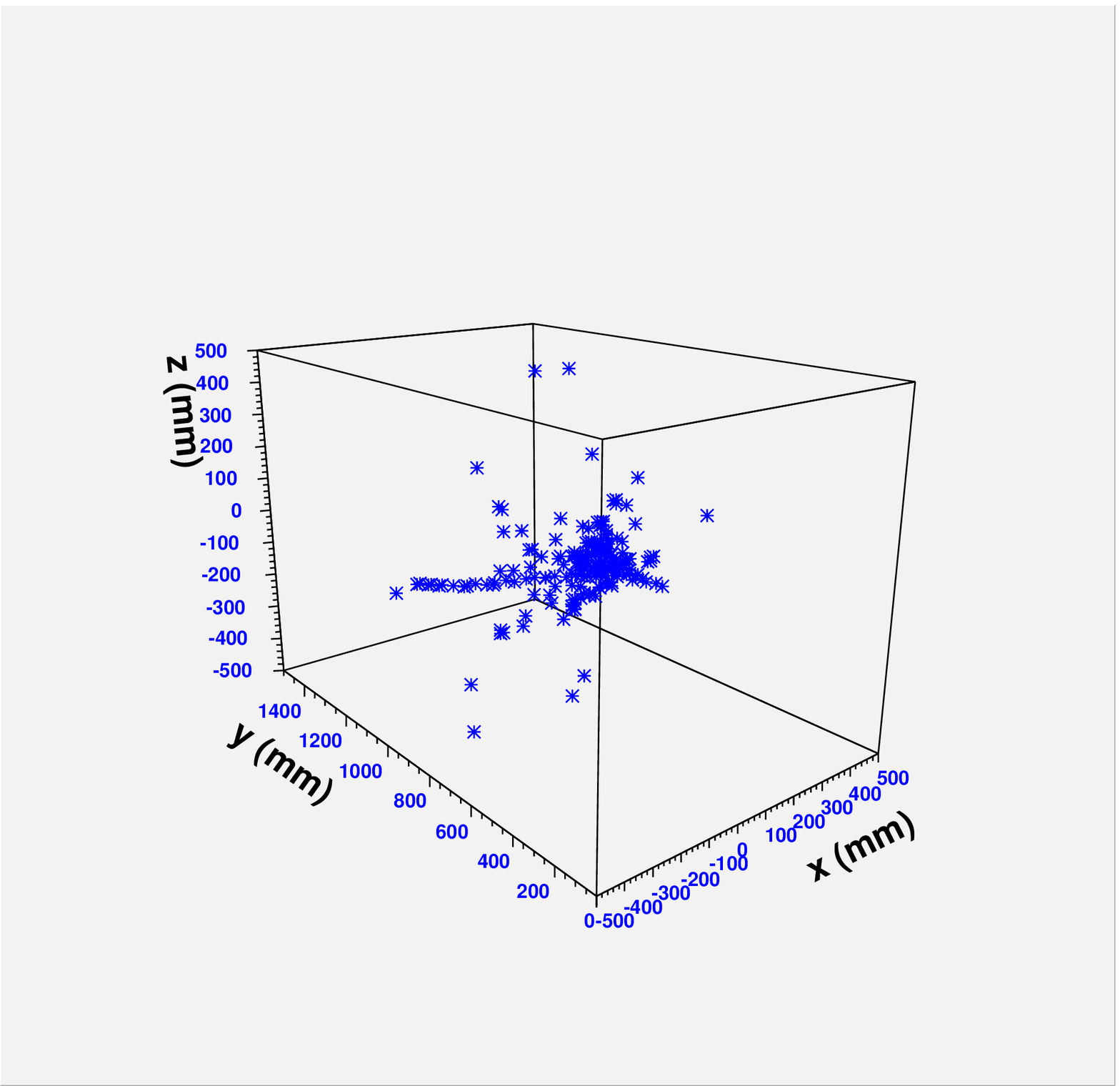,width=3.4cm}}&
\mbox{\epsfig{file=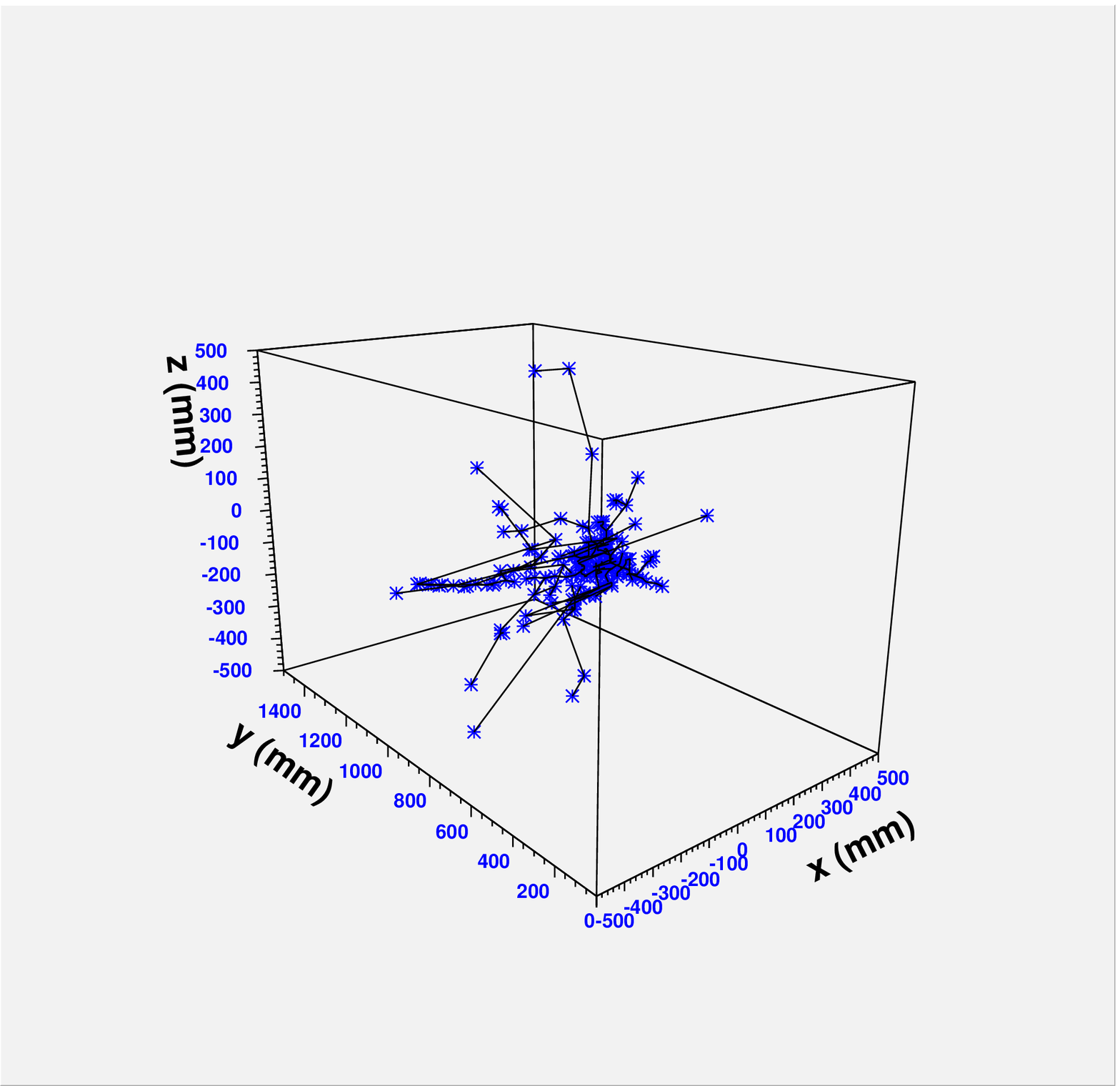,width=3.4cm}}&
\mbox{\epsfig{file=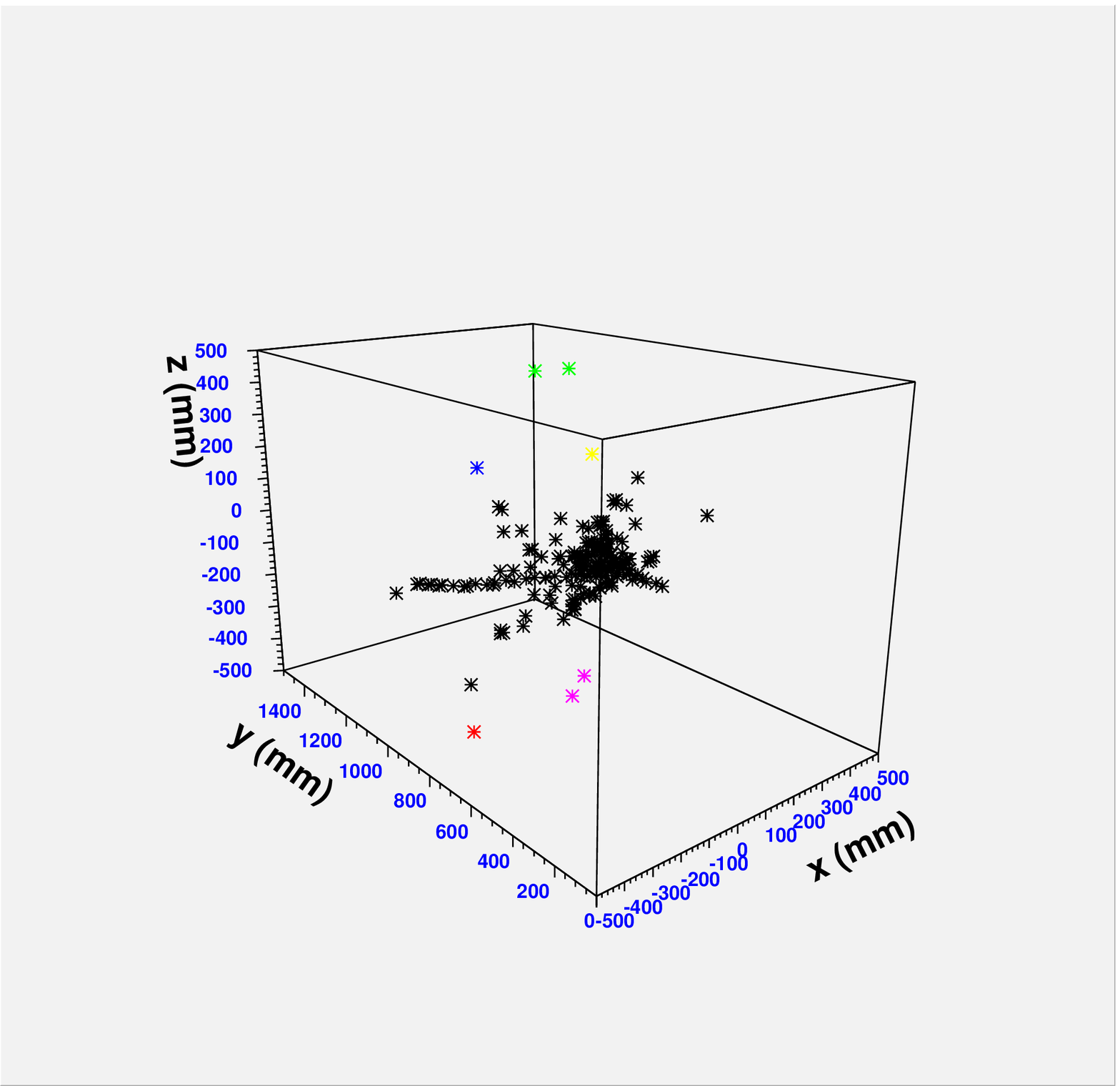,width=3.4cm}} \vspace{-10pt}\\
\end{tabular}
\end{center}
\caption{Example, clustering a single hadronic shower.}
\label{fig:two}
\end{figure}

A clustering algorithm based on minimal spanning trees has been developed.
It can operate standalone or perform preclustering before a sophisticated 
energy-flow algorithm is applied [\ref{ainsley}]. 
Its operation is divided into three consecutive steps.
First an appropriate metric, not necessarily euclidean, should be defined. 
Then the corresponding minimal spanning tree is constructed using 
Prim's algorithm~[\ref{prim}]. 
The final step is to perform single linkage cluster analysis {\em i.e.} go through 
the tree and cut the branches with length above a proximity bound that the  
nodes belonging to the same cluster must obey. 
The algorithm is an $\cal{O}$$(N^2)$ loop, where $N$ is the number of nodes.
Also it should be emphasized that after defining an appropriate metric 
for the problem the rest of the algorithm has no dependency on detector geometry 
since only the metric deals with this. 
First tests of the algorithm with single and multiparticle events show satisfactory 
performance. A simple example is depicted in Fig~\ref{fig:two}.

\section{Summary}

We have discussed a top-down approach to calorimeter clustering 
based on minimal spanning trees, highlighting in brief their theoretical 
background and implementation in a clustering algorithm.

\end{document}